\documentstyle[12pt]{article}
\textwidth
16.4cm
\oddsidemargin
2.5cm
\advance\oddsidemargin by
-1in
\evensidemargin
0.0cm
\advance\evensidemargin
by
-1in
\marginparwidth
1.9cm
\marginparsep
0.4cm
\marginparpush
0.4cm
\topmargin
-1.5cm
\advance\topmargin
by
-0.0in
\textheight
23.5cm
\makeindex

\newcommand\noi{\noindent}
\newcommand\beq{\begin{equation}}
\newcommand\eeq{\end{equation}}
\newcommand\beqn{\begin{eqnarray}}
\newcommand\eeqn{\end{eqnarray}}

\newcommand{\doublespace}
{
\renewcommand{\baselinestretch}
{1.6}
\large\normalsize}

\begin{document}
\vspace*{1cm}
\hspace*{9cm}{\Large MPI H-V18-1997}

\medskip

\hspace*{9.0cm}{\Large DFTT-24/97}
\doublespace

\vspace*{3cm}

\centerline{\Large \bf Unitarity effects in
DIS}
\medskip

\vspace{.5cm}
\begin{center}
 {\large Boris~Kopeliovich$^{1,2}$, Bogdan
Povh$^{1}$
and Enrico
Predazzi$^{3}$}

\vspace{0.3cm}

$^{1}$ {\sl Max-Planck Institut
f\"ur
Kernphysik,
Postfach
103980, 69029 Heidelberg,
Germany}\\

$^{2}${\sl Joint
Institute
for Nuclear Research, Dubna,
141980
Moscow Region,
Russia}\\

$^3${\sl Universit\`a di
Torino
and INFN, Sezione di Torino, I-10125, Torino,
Italy}

\end{center}

\vspace{1cm}
\begin{abstract}
We argue that diffractive DIS, dominated by soft interactions,
is probably the unique process which allows us to observe 
unitarity effects in DIS.
Guided by a close analogy between the diffractive dissociation of
a highly virtual photon and the elastic scattering of hadrons
we propose a specific procedure to analyse the data in order to detect
the onset of the unitarity limit.
Lacking appropriate data, we use the predictions of a realistic model
as an input for our analysis, to demonstrate that the output unitarity 
signal is sufficiently large to be detectable.

\end{abstract}

\bigskip
\doublespace
\newpage
\noi
{\bf Introduction}
\medskip

Diffractive phenomena, or large-rapidity-gap 
events in deep-inelastic scattering (DIS) 
have become one of the central
issues of HERA physics. Diffraction, according
to its optical analogy, is usually 
associated with shadowing and is viewed as a 
domain of soft hadronic interactions. 
Therefore, one could naively expect
that diffraction in DIS vanishes at high $Q^2$; this
is not the case according to recent experimental 
HERA data \cite{h1,zeus}. 
We begin by commenting
on the similarities and differences 
between diffraction in hadronic interactions and
in DIS. Since manifestations of saturation of the unitarity limit
have already been found long ago in elastic hadronic scattering,
one may hope to detect similar effects in diffractive 
DIS.

The main objective of this
paper is to clarify if and how one could observe unitarity effects
in DIS. In particular, can the data
reveal the onset of unitarity or whether the unitarity 
limit has already been reached in the diffractive cross section
at HERA energies?  In DIS,
a substitute for the elastic cross section has to be found which
enables one to perform an analysis analogous to that of hadronic reactions.
We give a specific prescription of how to analyse the diffractive
DIS data in order to identify unitarity effects. Modelling the
diffraction cross section within a realistic model we demonstrate
that the expected signal of unitarity limit is 
nearly as large as in hadronic elastic scattering.

\medskip

\noi{\bf Diffraction in DIS}
\medskip

DIS is traditionally viewed as a hard probe of the
parton distribution in hadrons. However, looking at it in the
proton rest frame, which is better suited to discuss diffraction 
in DIS, one easily realises that the interaction of a virtual photon
has a substantial soft contribution even at high $Q^2$.
This argument was first put forward by Bjorken and Kogut
\cite{bjorken} with phase space motivations.
It was further demonstrated within perturbative QCD
\cite{nz91} that the soft contribution 
to the total DIS cross section scales with 
$Q^2$ and is not a higher twist effect.
The reason why the soft component does not vanish at
high $Q^2$
can be made plausible in the following way \cite{kp96}.
Soft hadronic
fluctuations of a highly virtual photon are expected to
be rare; they are suppressed
by a factor $\sim 1/Q^2$ (for transversely polarized photons 
\cite{nz91}) as compared to 
hard fluctuations. But soft fluctuations have a large cross section,
of the order of a typical hadronic cross section. On the other hand,
hard fluctuations, due
to color screening, have a small size and, therefore, a tiny
interaction cross section ($\sim 1/Q^2$).  The net result is
that the two contributions to the DIS cross section,
soft and hard, end up having the same leading twist behaviour.
This conclusion is supported by the experimental observation 
of a rather weak $Q^2$-dependence in nuclear shadowing \cite{kp96}.

In the case of diffractive scattering, the equilibrium
between soft and hard components breaks down. The
diffractive cross section is still proportional to the same
probability of finding the appropriate fluctuation (hard or
soft) within the photon; it is, however, not
proportional to the total cross section, but to
its square (see \cite{zkl81}).  As a consequence, hard
fluctuations get an extra factor $1/Q^2$ and become a
higher twist effect in diffraction while, on the contrary, 
soft fluctuations in diffractive DIS remain leading twist.
 

These considerations are, obviously, very qualitative.
The distinction between soft and hard components is quite
conventional; in addition, there is an intermediate region
of semi-hard fluctuations (whose size is $Q^2$-independent 
although quite small). The question then is whether the
above conclusions are confirmed by the data.  The answer
is yes, and comes from nuclear shadowing in DIS which
is closely related to diffraction.  The same size, which
dominates the photon fluctuations in the diffraction
dissociation on a proton, is also responsible for the
first order nuclear corrections.  The size of nuclear shadowing
observed in heavy nuclei at low $x$ \cite{nmc} requires a
cross section of about $12\ mb$, which corresponds to a $q\bar q$ 
separation of nearly $0.6\div 0.8\ fm$; this proves that, indeed,
soft diffraction is the dominant contribution to
diffractive DIS.

From the H1 \cite{h1} and ZEUS \cite{zeus} experiments at
HERA, we know that the fraction of photon diffraction in DIS
cross section is about $10\ \%$.  At first sight one may
think that this is the analogue of what one finds in hadronic
interactions, where about the same fraction of total cross
section comes from single diffraction.
But this is not so. In the virtual photon interaction, diffraction
corresponds to both, the elastic and the inelastic diffractive
scattering of the fluctuations. 
For instance, in the photon interaction, the diffractive cross section
on the black disk reaches $ 50\ \%$  of the total
cross section \cite{nzz}. 
In hadronic collisions, the inelastic diffraction
would vanish, while the elastic cross section becomes
$50\ \%$ of the total cross section. 
For a meaningful comparison with hadronic
interactions we have to look at the fraction of the total
cross section taken by elastic and single inelastic
diffraction {\it i.e,} about $25\ \%$ in $\pi p$ and $pp$
interactions; this is substantially larger than in
diffractive DIS. The reason is clear, while the contribution
of hard interaction to diffraction vanishes, it still
provides more than half of the total DIS cross
section. 

After realising that the diffraction dissociation of a highly
virtual photon is predominantly a soft process and that the elastic
scattering of the photon fluctuations play an important role,
one may expect that unitarity effects in diffractive DIS
could be about as important as in hadronic reactions. 
Before addressing the question of the unitarity in DIS,
let us briefly recall
some relevant points concerning how the same question
has been approached in soft hadronic physics.
\medskip

\noi{\bf Unitarity in hadronic reactions}
\medskip

It is known that a growth of total cross sections as a power
of energy would ultimately lead to a violation of unitarity;
the latter, in fact, restricts the rate of growth of total
cross sections to $ln^2(1/x)$ according to Froissart's
theorem.  In hadronic interactions, the data are commonly
reproduced either using unitarity non violating logarithmic
forms or a power behaviour whose growth is so gentle
($\sim
(s/s_0)^{0.1}$) that a noticeable violation of the Froissart-Martin
bound is
expected only at energies far beyond the range of present
accelerators \cite{landshoff}. By contrast, a much steeper
growth with energy of the cross section for interaction
of highly virtual photons was discovered at HERA,
$\sigma_{tot}^{\gamma^*N} \propto (1/x)^{\Delta}$, where the
power $\Delta$ reaches much larger
values, $0.3 \div 0.4$ at high $Q^2$.  This fact led to the
widespread expectation of an earlier onset of unitarity in
DIS as compared with soft hadronic interactions, which is
supposed to show up as a slow down of the observed
growth of $\sigma_{tot}^{\gamma^*N}$ at large $1/x$.
It is, however, not obvious {\it how} we could claim to
observe such a slow down, since we do
not know any reliable baseline to search for a deviation
from, {\it i.e.} we do not have any rigorous
theoretical prediction for this $x$-dependence 
to compare with.  The power behaviour
$(1/x)^{\Delta}$ would correspond to the Pomeron if it were
a Regge-pole, but it is not, since $\Delta$ grows with $Q^2$
violating factorization.  Thus, this behaviour has to be regarded merely as
a convenient parameterization not dictated by any rigorous
theory especially in the HERA energy range.  For instance,
the double-leading-log approximation \cite{nzz1} predicts
that the effective power $\Delta(x) \to 0$ in the limit $x\to 0$.

In addition, locally, unitarity sets restrictions only on the
magnitude of the imaginary part of the partial elastic
amplitude, not on its variation with energy.  No detectable
unitarity effect can be unambiguously ascertained if the
amplitude is much smaller than one (like in $\gamma^*p$
interaction) irrespective of its actual energy-dependence.
The observed steep growth of $\sigma_{tot}^{\gamma^*N}$ at
low $x$ is supposed to originate from the dominant photon
fluctuations, which have a size $\sim 1/Q^2$ \cite{nzz1}. As
already mentioned, however, the cross section for these
fluctuations is tiny ($\sim 1/Q^2$) and cannot
cause any problem with unitarity, independently of its energy
dependence.  On the other hand, the soft component of DIS
corresponding to large size hadronic fluctuations may need
some unitarity corrections, even though it is supposed to
have about the same weak energy dependence as in hadronic
interactions.  We believe that it is hopeless to observe
those corrections in the total DIS cross section.  Even in
hadronic interaction, $\sigma_{tot}^{pp}$ shows no deviation
from an effective power energy dependence up to the highest
available energies \cite{landshoff}. 
Unitarity corrections were nevertheless detected and shown to be important
in hadronic interactions in the analysis of elastic
scattering data over the ISR energy range \cite{amaldi}.

Let us write the hadronic (say $pp$ for
definiteness)
differential cross section as ${{d\sigma}\over {dp_T^2}}
=
|R(s,p_T^2)+ i A(s,p_T^2)|^2$ where $R(s,p_T^2)$
and
$A(s,p_T^2)$ are the real and the imaginary parts of
the
amplitude and $\vec p_T$ is the proton transverse
momentum.
At high energies we know the real part to contribute a
relatively
small (say $\leq 1\%$) fraction of the differential
elastic
cross section at small $p_T^2 \leq 0.5 GeV^2$, where
about
$99 \%$ of the events are concentrated.  Neglecting
$R(s,p_T^2)$ one can simply invert the above equation
to
determine $A(s,p_T^2) \approx \sqrt{d\sigma/dp_T^2}$
and
apply a Fourier transform to obtain the approximate
{\it
partial wave} amplitude $F(s,b)$ in the
impact-parameter
representation, for which the unitarity condition reads
$$0
\leq G_{in}(s,b) = 2 Im F(s,b) - |F(s,b)|^2 \leq 1$$
(our
notation differs from that of \cite{amaldi} by a factor of $i$
so
that what is their $Real$ part becomes our $Imaginary$
one
and vice versa).  In the literature, $F(s,b)$
and
$G_{in}(s,b)$ are also known as the {\it profile
function}
and the {\it inelastic overlap function} respectively. As
a
consequence of the above condition, 
$0 \leq Im F(s, b) \leq
1$.

As found in \cite{amaldi} for the ISR $pp$
and $\bar p p$ data the {\it
normalization of $G_{in}$ at $b=0$ is 
so close to unity as to
essentially saturate unitarity}. Therefore, an
extrapolation
to higher energies of the supercritical Pomeron
form
$(s/s_0)^{\Delta}$ would violate unitarity at $b=0$
already
at the $Sp\bar pS$
energies.

The important result of \cite{amaldi}, however,
is that if one takes the difference $\Delta G_{in}(s,b)$
(evaluated from the data) of $G_{in}(s,b)$ calculated at the
{\it lowest} ($23\ GeV$) and at the {\it highest} ($63\
GeV$) ISR energies, it is almost entirely concentrated on
the periphery at $b \sim 1\ fm$ (see Fig.8 of \cite{amaldi}).
This is the crucial observation which we propose as a guide on
how to analyze the DIS diffraction data so as to make
evident the effects of unitarity.
\medskip

\noi{\bf Procedure of Analysis}
\medskip

The procedure we suggest in order to detect
possible unitarity effects from the data on diffractive DIS
is essentially the same as we have just described for
hadronic reactions.  We treat the various channels of
diffractive dissociation of a virtual photon as elastic
scattering of hadronic fluctuations of the
photon in its eigenstate basis (e.g. $|q\bar q\rangle,\
|q\bar qg\rangle$, etc. with definite transverse separations).

Given the diffractive dissociation cross section
$d\sigma^{\gamma^* p}_{dd}(x,Q^2)/dp_T^2\ dM^2$,
we sum over all events with different $Q^2$s
in order to increase the statistics. This is allowed because
there is no substantial $Q^2$ dependence in
the diffractive cross section (as expected since
diffraction is dominated by soft interactions). Then, we sum
over the final states by integrating over $M^2$.

\beq
\frac{d\sigma^{\gamma^* p}_{dd}}{dp_T^2}(x)
=
\int\limits_{M^2_{min}}^{M^2_{max}}
\frac{d\sigma^{\gamma^* p}_{dd}}{dp_T^2\
dM^2}dM^2
\label{10}
\eeq

\noi
One should confine oneself to the mass interval where a
$q\bar q$ component of the photon dominates, because it is
the softest one, and one should cut off the low-mass
resonance production, which is a hard process. As a
suggestion, one can use $M^2_{min} \approx 5\ GeV^2$.
The region $M^2 \gg Q^2$ is known to be dominated by the
triple-Pomeron term, which corresponds to the higher Fock
components in the photon, $|q\bar qg\rangle$, $|q\bar
q2g\rangle$, etc.  These fluctuations are expected to have
a smaller size because we know from lattice calculations the
shortness of the gluon correlation radius 
($\sim 0.2\div 0.3\ fm$).  To suppress this contribution one should
restrict the mass interval to $M^2 \sim Q^2$; typically, one can
take $M^2_{max}$ equal to a few units of $Q^2$.  The exact value
does not seem so important because the high
mass tail $d\sigma_{dd}/dM^2 \propto 1/M^2$ does not
contribute much. Note that these higher Fock components, or
the triple-Pomeron contribution to diffraction is
just another aspect of the gluon-gluon fusion mechanism
which is a widely recognised manifestation of unitarity.

Next, we take the square-root of the differential
cross
section (\ref{10}) and perform a Fourier transformation to
the
impact-parameter
representation

\beq
F(b,x) = \int d^2p_T\ e^{i\vec p_T\vec
b}
\sqrt{\frac{d\sigma_{dd}^{\gamma^*
p}}{dp_T^2}}
\label{11}
\eeq

Contrary to the hadronic case, we {\it cannot set
any
unitarity restriction} on the value of $F(b,x)$, since
it
includes factors (the fine structure constant, etc.),
which
cannot be given in a model-independent way.  For this
reason
we do not care about the normalization of $F(b,x)$.  The
way
we suggest to observe unitarity effects is by studying
the
$x$-dependence of $F(b,x)$.  In strict analogy with the
hadronic case, we expect only {\it a slow growth of $F(b,x)$ with
$1/x$ at small $b$ and the main increase  at large} $b$.  To check
this conjecture, we define the slope of the $x$-dependence in
the normalization-free way

\beq
\Delta_{eff}(b) =
\frac{d\ln[F(b,x)]}
{d\ln(1/x)}
\label{12}
\eeq

In the case of elastic hadron scattering
$\Delta_{eff}(b)$
is $b$-independent, the unitarity corrections are
neglected, and is related to the Pomeron intercept,
$\Delta_{eff} = \alpha_P(0) - 1$.  Since any unitarity
correction is expected to slow down the rate of growth of
$F(b,x)$ with $1/x$ for central collisions,
$\Delta_{eff}(b)$
should be smaller at small $b$ than at large $b$.
This is the $\it signature$ of unitarity effects we expect.
\medskip

\noi{\bf Theoretical experiment}
\medskip

Given that for the time being we are not aware of any data 
that allow us to perform this
analysis (these data may, however, well have already been collected), 
the rest
of this paper will be devoted to check our prediction of unitarity
effects with a model calculation. It is not only a problem
of statistics (which such an analysis needs) which one
may have to worry about, but also of other
complications which are absent in elastic hadronic scattering
and which could well wash away the signal in the diffractive
dissociation of a highly virtual photon. Firstly, there
is a background of hard and semi-hard photon fluctuations,
from which we do not expect any unitarity effect. In addition, it 
is not obvious how the interplay between many channels in
the diffractive interaction of the photon's fluctuations
may affect unitarity.

To check the possibility of observing an onset of unitarity
we suggest a theoretical experiment, {\it i.e.}
we calculate the differential cross section
$d\sigma^{\gamma^* p}_{dd}/dp_T^2$ in a crude but realistic
model
based on pQCD, and then we use the result as an input
for our analysis as if it were experimental data.

In the light-cone formalism the photon diffractive dissociation 
cross section due to $q\bar q$ fluctuations
can be written as \cite{nz91},
\beq
\frac{d\sigma^{\gamma^* p}_{dd}(x)}{dp_T^2}
=
\int\limits_0^1\
d\alpha
W_{\gamma^*}(Q^2,\alpha,\rho)\
\int
d\rho^2\
\frac{d\sigma^{(q\bar
q)p}_{el}(\rho,x)}{dp_T^2}
\approx
K\
\int\limits_{1/Q^2}^{1/\mu^2}
\frac{d\rho^2}{\rho^4}\
\frac{d\sigma^{(q\bar q)
p}_{el}(\rho,x)}{dp_T^2}
\label{13}
\eeq
\noi
Here $\rho$ is the transverse separation in the $q\bar
q$
fluctuation, $\alpha$ is the fraction of the photon light-cone 
momentum carried by the quark and $W_{\gamma^*}(Q^2,\alpha,\rho)$ 
is the probability of such a $q\bar q$ fluctuation in the photon. 
The integration over $\alpha$ has been performed using the fact \cite{nz94} 
that the function
$\rho^4\int\limits_0^1
d\alpha
W_{\gamma^*}(Q^2,\alpha,\rho)$ is approximately constant in
the interval $1/Q^2 < \rho^2 < 1/\mu^2$, where $\mu$ is a
hadronic scale mass parameter, of the order of $\Lambda_{QCD}$.
K is a normalization factor, irrelevant for further considerations.

The elastic differential cross
section
$d\sigma^{(q\bar q) p}_{el}(\rho,x)/dp_T^2$ for the
$q\bar
q$ pair with separation
$\rho$
can be estimated using a factorized dipole partial-wave amplitude
(a bare Pomeron to be further unitarized), of the form,
\beq
f^{(q\bar q)p}(b,\rho,x)
=
\frac{i}{2}\
\sigma(\rho)\
\left(\frac{1}{x}\right)^{\Delta_0}\
\exp\left[-\frac{b^2}{2B(\rho,x)}\right]\
,
\label{13a}
\eeq
(where we neglect the small real part) which depends on
$b$ (the relative impact parameter between the center of gravity of
the $q\bar q$ pair and the proton, on the transverse separation
$\rho$ inside the $q\bar q$ pair and on $x$. The
universal dipole cross section $\sigma(\rho)$ \cite{zkl81}, 
which vanishes $\propto \rho^2$ at small $\rho$, was estimated in
\cite{kz91}. The exponent $\Delta_0 \approx 0.1$, has a value typical
for soft hadronic interactions, due to the fact that diffraction is
soft dominated (this is confirmed by the analyses \cite{h1,zeus}
of the experimental data).
The slope parameter $B(\rho,x)$ is a smooth function
of $\rho$ and $x$. Its actual value is not so important for
detecting unitarity effects; for simplicity, we assume
it to be a constant, $B(\rho,x)\approx 5\ GeV^{-2}$ \cite{zeus}.

The central point of this procedure is unitarization of
the
input partial amplitude (the bare Pomeron)
(\ref{13a}).
It may exceed unity
and
violate unitarity at small $x$ (mainly at large $\rho$
and
small
$b$).
We use the popular eikonal form for
the
unitarized amplitude \cite{dklt}, $Im\
F^{(q\bar
q)p}(b,\rho,x) = 1 -
\exp\left[-
Im\ f^{(q\bar q)p}(b,\rho,x)\right]$, which is
exact in our case since a $q\bar q$
pair
with a fixed separation $\rho$ is an eigenstate of
the interaction,
{\it i.e.} no off diagonal intermediate states are
possible.
A nice property of the eikonalized amplitude is that it
guarantees
that unitarity is not violated, $Im\ F(b) \leq 1$. Going to lower
and
lower $x$ values, we expect it first to stop growing at small
$b$,
then this effect spreads to larger $b$'s
\cite{dklt}.
This is exactly the signature of unitarity limit we are proposing to
observe.

Once we have the partial wave amplitude, we can
Fourier transform it to calculate the differential elastic
cross section
\beq
\frac{d\sigma^{(q\bar q) p}_{el}(\rho,x)}{dp_T^2}
=
\frac{1}{4\pi}\ \left|\int d^2b\ e^{i\vec p_T\vec
b}\
F^{(q\bar
q)p}(b,\rho,x)\right|^2
\label{14}
\eeq
This differential cross section explicitly obeys the
unitarity restrictions. However, it should be averaged over
$\rho$ using (\ref{13}) to produce an observable
diffractive dissociation cross section, summed over final
state. Doing this
we arrive at the goal of these
calculation,
the differential cross section of diffractive
dissociation
which can be used until the data become
available
as an input
to
eqs.~(\ref{11})-(\ref{12}).
The result for the $b$-dependent
$\Delta_{eff}(b)$
is shown by the solid curve in Fig.~1. It
indeed
demonstrates what we had
anticipated, a suppressed value of $\Delta_{eff}$, 
{\it i.e.} a substantially less steep
$x$-dependence, at small $b$. At $b>1\ fm$ the result
of the Fourier transform is extremely sensitive to the details
of the model, which can even  cause oscillations of
the partial amplitude. However the amplitude itself is
tiny at these impact parameters, and we do not expect
experimental data to provide a reasonable accuracy in
this region. Notice that ZEUS and H1 spectrometers are
blind at $p_T^2 < 0.1\ GeV^2$ \cite{columbia}, however,
we checked that this may affect $\Delta_{eff}$ only at
$b>1\ fm$. The signal of unitarity limit predicted
at $b<1\ fm$ is unambiguous.

\begin{figure}[tbh]
\includegraphics{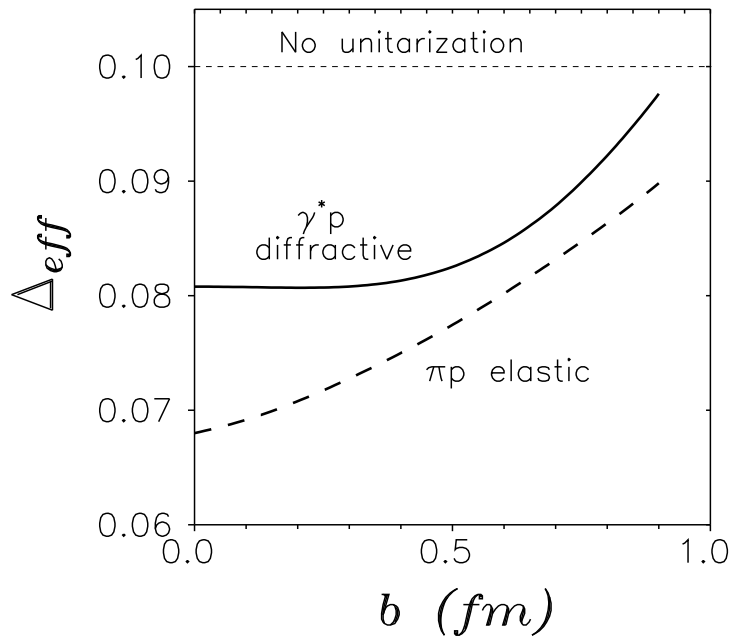}
\begin{center}
\vspace{9cm}
\parbox{13cm}
 {\caption[Delta]
 {Predicted $b$-dependence
of the effective exponent $\Delta_{eff}(b)$ defined
in (\ref{12}) for diffractive DIS (solid curve)
and for elastic pion-nucleon scattering (dashed curve).
The dotted line $\Delta_{eff}(b)=\Delta_0$
corresponds to the same analyses, but without unitarity
corrections}
\label{fig1}}
\end{center}
\end{figure}

Notice that if the non unitarized amplitude (\ref{13a})
is used as an input for the analysis, the resulting 
exponent $\Delta_{eff}(b)$ is $b$-independent and is
equal to $\Delta_0$.

To obtain the feeling of how the suppression of
$\Delta_{eff}(b)$
at small $b$ in diffractive DIS looks compared with that in
hadronic
interaction we have performed the same analysis for
pion-proton
elastic scattering. The photon wave function is replaced
with
the pion one which we model by a
Gaussian
with the appropriate charge radius and with a
$\delta(\alpha-1/2)$
distribution over $\alpha$. We use the same
universal
$\rho$-dependent amplitude (\ref{13a}), but the
larger slope parameter $B \approx 8\
GeV^{-2}$, corresponding to the experimentally observed slope
in the pion-proton elastic scattering.
The result for $\Delta_{eff}(b)$ is shown by the dashed curve in
Fig.~1.
Amazingly, the indication is that the unitarity limit effects in
diffractive
dissociation of a highly virtual photon are not much smaller than
in
pion-proton elastic scattering, in spite of the many corrections which
could
have diminished the effect.

{\it Summarizing}, we suggest that effects of unitarity
in DIS can indeed be detected in the diffractive DIS data.
We believe that adapting the same procedure of analysis
which has been proved effective in elastic hadron-hadron scattering
\cite{amaldi} is probably the unique way 
to pin down unitarity effects in diffractive data and
we have given a detailed recipe on how to proceed in this analysis.
Lacking appropriate data, we have demonstrated that the
procedure works quite effectively within a specific
but, we believe, realistic model. 

An experimental verification of the impact parameter dependence of the 
diffractive cross section would be
a clean manifestation of the dominance of the 
soft interaction in these processes.

\end{document}